\documentclass[twoside,fleqn]{article}
\usepackage{espcrc2}
\usepackage[dvips]{epsfig}
\newcommand{\be}{\begin{equation}}
\newcommand{\ee}{\end{equation}}
\newcommand{\bea}{\begin{eqnarray}}
\newcommand{\eea}{\end{eqnarray}}
\newcommand{\bean}{\begin{eqnarray*}}
\newcommand{\eean}{\end{eqnarray*}}
\newcommand{\dts}[2]{#1_{{\scriptsize \mbox{#2}}}}

\newcommand{\cl}[1]{\mathcal{#1}}

\newcommand{\Real}{\mbox{Re\,}}
\newcommand{\real}[1]{\mbox{Re}\left(#1\right)}
\newcommand{\Imag}{\mbox{Im\,}}

\newcommand{\dt}{\mbox{d}}
\newcommand{\Exp}{\mbox{e}}
\newcommand{\mv}[1]{\langle#1\rangle}

\newcommand{\half}{{\textstyle{\frac{1}{2}}}}
\newcommand{\hf}[1]{#1_{x}}
\newcommand{\hfup}[2]{#1_{x+\hat{#2}}}

\newcommand{\gf}[2]{#1_{x,#2}}

\newcommand{\pln}[2]{#1_{{\scriptsize \mbox{p}};#2}}
\newcommand{\lkop}[3]{#1_{#2;#3}}
\title{
{
\vspace{-4cm} \normalsize \hfill
\parbox{30mm}{MS-TPI-96-11\\hep-lat/9607048}
}\\[30mm]
Scaling in the two-dimensional U(1)--Higgs model}
\author{
Hermann Dilger and Jochen Heitger%
\thanks{Presenter at the conference LATTICE~96, St.~Louis.} \smallskip\\
Institut f\"ur Theor. Physik I, Universit\"at M\"unster,
Wilhelm-Klemm-Str.~9, D-48149 M\"unster, Germany}
\begin{document}
\makeatletter
\long\def\@maketablecaption#1#2{\vskip 10mm #1. #2\par}
\makeatother
\begin{abstract}
We study the continuum limit of the $2D$ U(1)--Higgs model with variable 
scalar field length, which is qualitatively different from the fixed 
length case.
Our simulations concentrate on the scaling behaviour of the topological 
susceptibility, and an instanton-induced confinement mechanism of 
fractional charges is numerically confirmed.
\end{abstract}
\maketitle
%
%
\section{Introduction}
The $2D$ abelian Higgs model shares prominent features of the 
SU(2)--Higgs sector of the Standard Model related to baryon number 
violation.
Whereas detailed studies of the model with various methods are available 
\cite{JKS79IM94,GIM87}, it is not well understood within the euclidean 
lattice approach, above all for variable scalar field length.
We examine the continuum limit in this case and, in particular, 
investigate the scaling behaviour of the topological susceptibility.
\section{Simulation of the lattice model}
On a two-dimensional lattice $\Lambda$ (with spacing $a$, extensions 
$L_{\mu}$, and unit vectors $\hat{\mu}$, $\mu=1,2$) the action is 
given by $S=\dts{S}{g}+S_{\phi}$,
\bea
\dts{S}{g}
& = &\beta\sum_{x\in\Lambda}\Big(1-\Real\pln{U}{x}\Big)
\label{WilAct}\\
S_{\phi}
& = &\sum_{x\in\Lambda}\Bigg\{-2\kappa\sum_{\mu=1}^{2}
     \real{\hfup{\varphi}{\mu}^*\gf{U}{\mu}\hf{\varphi}}\nonumber\\
&   &+|\hf{\varphi}|^2+\lambda\Big(|\hf{\varphi}|^2-1\Big)^2\Bigg\}\,.
\label{ScaAct}
\eea
$\pln{U}{x}\equiv\Exp^{i F_x}$ is the Wilson plaquette.
The gauge fields $\gf{A}{\mu}$ enter as phases of the links $\gf{U}{\mu}$, 
and the scalar field $\hf{\varphi}$ is decomposed as 
$\hf{\varphi}=\hf{\rho}\,\Exp^{i\hf{\omega}}$.

In the Monte Carlo simulation of this model a combination of
metropolis and overrelaxation algorithms (for the $\varphi$--field as
proposed in \cite{B95}) is applied.
Since these local algorithms generally do not manage to tunnel between 
different topological sectors, we use so-called instanton hits 
\cite{FS84D95}.
These are updates in the gauge sector by a global proposal of an instanton 
configuration
\be
\gf{A}{\mu}\,\rightarrow\,\gf{A}{\mu}\pm\Delta\gf{A}{\mu}
\label{IHit}
\ee
with $\Delta\gf{A}{\mu}$ carrying unit topological charge and being 
non-zero in a region of the instanton size.

We consider expectation values built up from the operators $\hf{\rho}$ 
(scalar length), the $\varphi$--links
\be
\lkop{L}{\varphi}{x\mu}^{\pm}\equiv\bigg\{
\begin{array}{c}
\Real(\hfup{\varphi}{\mu}^*\gf{U}{\mu}\hf{\varphi})\\
\Imag(\hfup{\varphi}{\mu}^*\gf{U}{\mu}\hf{\varphi})
\end{array}\,,
\label{PhiLnk}
\ee
and Wilson loops $W(R,T)$ of space-time extensions $R,T$. 
Particle masses in the Higgs ($m_H$) and vector ($m_W$) channels are
extracted from fits of $\hf{\rho}^2,\lkop{L}{\varphi}{x1}^+$-- and 
$F_x,\lkop{L}{\varphi}{x1}^-$--correlation functions, respectively.
\section{Lines of constant physics}
Let us mention some limiting cases of the model.
For $\kappa=0$ one arrives at pure gauge theory (PGT) with confinement in 
two dimensions.
$\lambda=\infty$ (fixed length case $|\hf{\varphi}|\equiv1$) and
$\beta=\infty$ is the $2D$ XY-model with its Kosterlitz-Thouless phase 
transition between a massive vortex phase ($\kappa\!<\!\kappa_c$) and a 
massless spin wave phase ($\kappa\!>\!\kappa_c$).
At finite $\beta$ this transition is expected to become a crossover 
\cite{JKS79IM94}.
For any fixed $\lambda$ and $\beta\rightarrow\infty$ the vector mass 
$am_W$ tends to zero, defining a continuum limit ($a\rightarrow 0$), but 
$am_H$ stays finite.
Ending up with infinite $m_H$ at $\beta=\infty$ for all (fixed)
$\lambda$--values reflects the freezing of the radial mode on large scales 
in the $2D$ $\phi^4_{n=2}$--theory \cite{P90}.
%
\begin{table*}[htb]
\begin{center}
\begin{tabular*}{\textwidth}{ccccccccccc}
\hline
  LCP & set & $\kappa$ & & $am_H$ & $am_W$ & $R_{HW} $ & $v_R$ &   
& $\dts{\chi}{top}\cdot 10^4$ & $\dts{\chi}{top}/m_H^2\cdot 10^4$ \\ 
\hline\hline
     & A & 0.2937  & & 0.882(4)  & 0.540(8) & 1.63(3) & 1.8736(1) &
& 0.105(4)                    & 0.14(1) \\
  L1 & B & 0.2607  & & 0.426(4)  & 0.258(5) & 1.65(4) & 1.8751(1) &
& 0.030(2)                    & 0.17(1) \\ 
     & C & 0.253   & & 0.221(1)  & 0.132(2) & 1.68(3) & 1.8772(1) &
& 0.011(1)                    & 0.21(2) \\
\hline 
     & A & 0.2858  & & 0.692(11) & 0.477(9) & 1.45(4) & 1.6144(2) &
& 1.62(2)                     & 3.4(2)  \\
  L2 & B & 0.25885 & & 0.330(5)  & 0.228(5) & 1.45(4) & 1.6170(2) &
& 0.53(1)                     & 4.9(2)  \\
     & C & 0.2525  & & 0.165(2)  & 0.112(3) & 1.47(4) & 1.6146(5) &
& 0.181(4)                    & 6.7(3)  \\
\hline
     & A & 0.2731  & & 0.446(11) & 0.514(8) & 0.87(2) & 1.2739(5) &
& 14.8(1)                     & 75(4)   \\
  L3 & B & 0.257   & & 0.234(8)  & 0.268(7) & 0.88(4) & 1.2757(5) &
& 4.12(3)                     & 75(6)   \\
     & C & 0.252   & & 0.121(2)  & 0.141(3) & 0.85(3) & 1.2746(5) &
& 1.21(1)                     & 83(3)   \\
\hline
\end{tabular*}
\parbox{\textwidth}{
\vspace{-0.75cm}
\caption{\label{LCPres1} \sl LCP-parameters
                             A: $\{\Lambda=16\times 16,\beta=10,
                             \lambda=0.01/0.013\}$,
                             B: $\{\Lambda=32\times 32,\beta=40,
                             \lambda=0.0025\}$ and
                             C: $\{\Lambda=64\times 64,\beta=160,
                             \lambda=0.000625\}$.
                             All mass errors come from a jackknife
                             analysis.}
}
\vspace{-0.75cm}
\end{center}
\end{table*}
%
%
\begin{figure}[htb]
\begin{center}
\epsfig{file=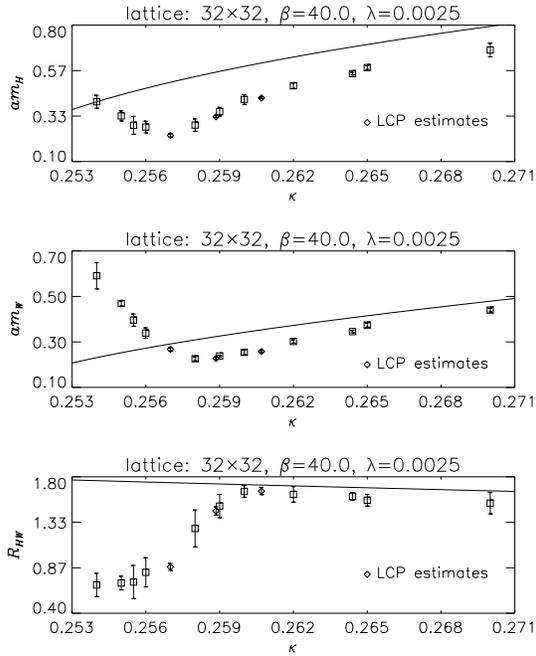,width=7.5cm}
\parbox{7.5cm}{
\vspace{-1.0cm}
\caption{\label{MassPlot} \sl Higgs and vector ($L_{\varphi}^-$) masses for 
                              fixed $\beta$ and $\lambda$ and their classical 
                              relations (solid lines).}
}
\vspace{-1.25cm}
\end{center}
\end{figure}
%

Figure~\ref{MassPlot} illustrates the typical dependence of the Higgs and 
vector masses on $\kappa$.
We find a change in the behaviour of the mass spectrum in addition to a
rapid breakdown of the topological susceptibility around the crossover  
$\kappa$--value $\bar{\kappa}$ defined at minimal $am_W(L_{\varphi}^-)$.
The $am_W$--estimates from $\varphi$--link ($L_{\varphi}^-$) and plaquette 
($F$) correlations are only consistent for $\kappa\!>\!\bar{\kappa}$.
For decreasing $\kappa\!<\!\bar{\kappa}$ the $F$--correlations weaken 
strongly and show a decreasing mass.
This corresponds to a massless photon in the PGT limit, which is, however, 
no physical degree of freedom.
The qualitatively different behaviour of the $L_{\varphi}^-$--correlations
in figure~\ref{MassPlot} is another striking analogy to the $4D$
model \cite{E87}, besides the fact that the $\kappa$--dependence of $am_H$ 
and $am_W$ is similar.  

Now we set up the lines of constant physics (LCPs) by the requirements
$am_H,am_W\!\rightarrow\!0$ at fixed scalar field VEV and mass ratio:
\be
v_R\equiv\sqrt{2\kappa}\,\mv{\rho}=\overline{v}
\quad\quad
R_{HW}\equiv\frac{m_H}{m_W}=\overline{R}\,.
\label{RenCond}
\ee
With a tuning of $\kappa$ this can be achieved by 
\mbox{$\beta\rightarrow\infty,\lambda\rightarrow 0$}, realized for large 
enough $\beta$ by
\be
\beta\rightarrow\infty\quad\quad\beta\lambda=\mbox{constant}\,.
\label{ContLim}
\ee
The simulated points in parameter space are collected in 
table~\ref{LCPres1}. 
One has $\beta\lambda\simeq 0.1$, and $\kappa$ was adjusted until the 
renormalization conditions (\ref{RenCond}) were simultaneously fulfilled 
within errors.

It has to be emphasized that the continuum limit (\ref{ContLim}), which 
amounts to send $\kappa\rightarrow\frac{1}{4}$ at the same time, see 
figure~\ref{ChiPlot}, should not be confused with the gaussian limit.
The crucial point is that the relation between the dimensionful bare 
continuum couplings $\lambda_0,e_0$ and the lattice parameters is 
$\lambda\propto a^2\lambda_0$ and $\beta=1/a^2e_0^2$.
Hence $\lambda\rightarrow 0$ at constant $\beta\lambda$ does not imply 
$\lambda_0\rightarrow 0$ for $a\rightarrow 0$.
\section{Topological susceptibility}
\begin{figure}[tb]
\begin{center}
\vspace{0.1cm}
\epsfig{file=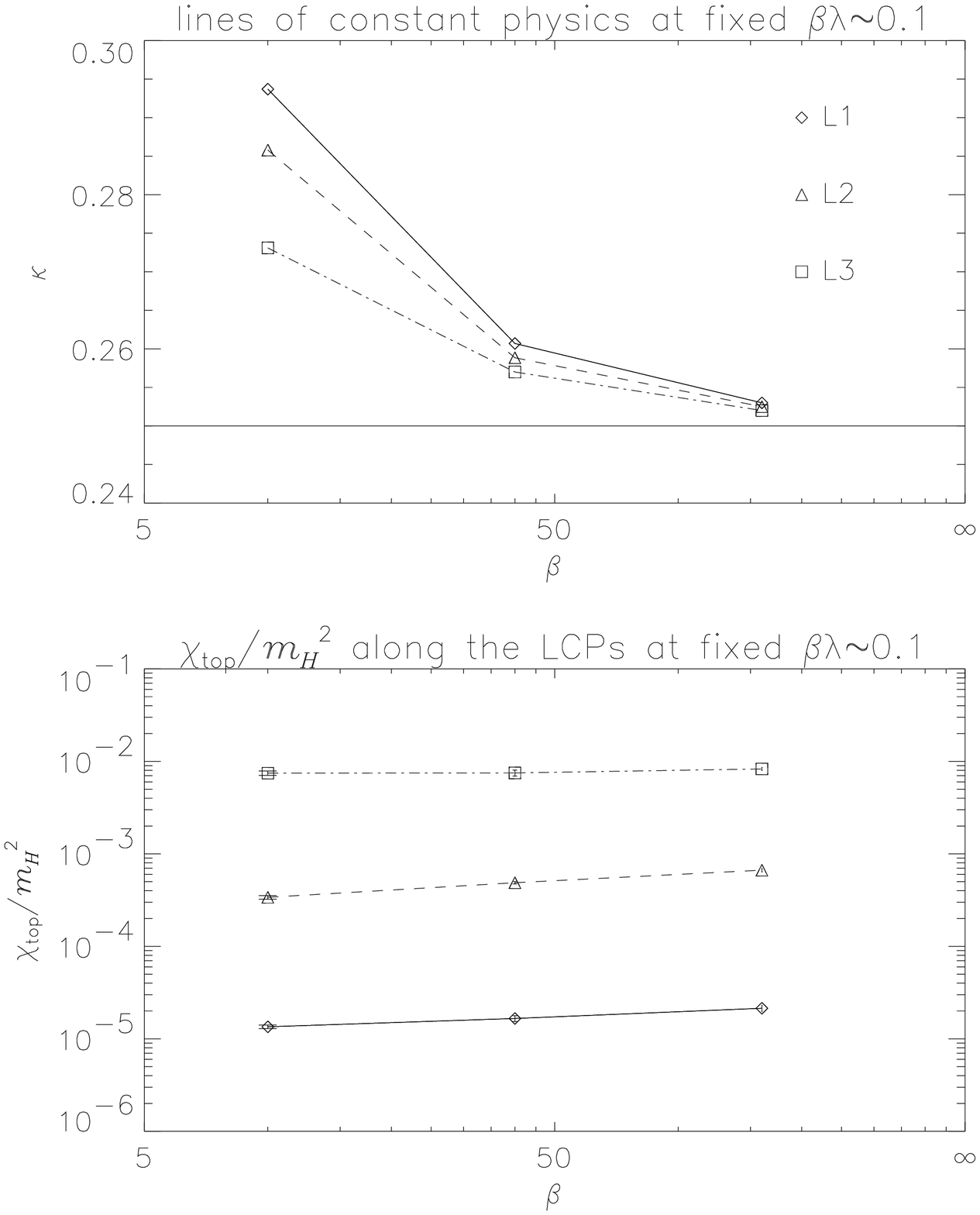,width=7.5cm}  
\parbox{7.5cm}{ 
\vspace{-1.0cm}
\caption{\label{ChiPlot} \sl Scaling of $\dts{\chi}{top}$ along the LCPs.}
}
\vspace{-1.25cm}                            
\end{center}
\end{figure}

We adopt the geometric definition of the topological charge, which in two
dimensions reads
\be
\dts{Q}{top}
\equiv\frac{e_0}{4\pi}\int\dt^2x\,\epsilon_{\mu\nu}F_{\mu\nu}(x) 
\,\longrightarrow\,\frac{1}{2\pi}\sum_{x\in\Lambda}F_x
\label{QTop}
\ee
and has only integer values. 
The topological susceptibility is 
$\dts{\chi}{top}\equiv\frac{1}{\Omega}\,\mv{\dts{Q}{top}^2}$,
$\Omega$: lattice volume, and has been measured on the LCPs leading to 
the results in table~\ref{LCPres1} and figure~\ref{ChiPlot}.
Significant finite volume effects are ruled out.
%
%
Within the chosen parameter sets $\dts{\chi}{top}$ varies by orders of 
magnitude, and a contraction of the $\kappa$--region, which is limited by 
a still measurable $\dts{\chi}{top}$ from above and by the line L3 from 
below, is seen. 
Note that this LCP already lies close to PGT, where the 
$\beta$--dependence $\dts{\chi}{top}\rightarrow\frac{1}{4\pi^2\beta}$
for $\beta\!\rightarrow\!\infty,\Omega\!\rightarrow\!\infty$ is known.
Except for L3, the scaling of the dimensionless ratio 
$\dts{\chi}{top}/m_H^2$ is rather poor.
 
Finally we look for confinement by instantons, suggested for this model 
in \cite{CDG77}.
Using
$\oint_{\partial\cl{A}}A_{\mu}\dt x_{\mu}=\int_{\cl{A}}\dt^2x\,F_{12}$
we obtain a unique lattice prescription for the Wilson loop with 
fractional test charge $q$ in the compact formulation:
\be         
W_q(R,T)=\Exp^{iq\sum_{x\in\cl{A}_{R,T}}F_x}\,\,\,
\cl{A}_{R,T}\in\Lambda:\,\mbox{area}\,.
\label{PLoop}
\ee
Since $F_x\!=\!e_0a^2F_{12}(x)$ for $a\rightarrow 0$, one requires
$F_x\!\in\![-\pi,\pi)$, so $2\pi$--ambiguities for $q\!=\!\half$ as for the 
standard form with $\gf{A}{\mu}$ are avoided.
The static potential 
$V_q=-\lim_{T\rightarrow\infty}\,\frac{1}{T}\,\ln W_q$
gets in the dilute instanton gas approximation a contribution
$\dts{\chi}{top}\{1-\cos(2\pi q/e_0)\}R$, which signals confinement for
non-integer $q/e_0$.
We take Polyakov loop correlations
$P_q(R)\equiv W_q(R,T)\,|_{T=L_2}$ and fit 
$V_q=-\frac{1}{q^2L_2}\,\ln P_q$ to a continuum Yukawa ansatz
\be
V_q(R)=\frac{e_R^2}{2m_s}\,\Big(1-\Exp^{-m_sR}\Big)
       \,+\,\alpha R\,.
\label{YukFit}
\ee
As exemplarily displayed in table~\ref{LCPres2} for $q=\half$ in L2,
lying just in the Higgs regime ($\kappa\!>\!\bar{\kappa}$), the meaning of 
the fit parameters $ae_R$ (renormalized gauge coupling, small corrections 
to $ae_0=1/\sqrt{\beta}$ expected), $am_s$ (screening mass, $\simeq am_W$) 
and $\alpha$ ($=2\dts{\chi}{top}/q^2$) is reproduced.
%
\begin{table}[htb]
\begin{center}
\vspace{-0.5cm}
\begin{tabular*}{7.5cm}{ccccc}
\hline
  set & & $am_s$ & $ae_R$ & $\alpha/8\cdot 10^4$ \\ 
\hline\hline
  A & & 0.436(2) & 0.3136(3) & 1.5(1)  \\
  B & & 0.209(2) & 0.1551(3) & 0.55(6) \\
  C & & 0.098(3) & 0.0769(5) & 0.20(5) \\
\hline
\end{tabular*}
\parbox{7.5cm}{
\vspace{-0.75cm}
\caption{\label{LCPres2} \sl Fit parameters of $V_{\frac{1}{2}}$ in L2.}
}
\vspace{-1.5cm}
\end{center}
\end{table}
%
\section{Discussion and outlook}
The continuum limit in the $2D$ U(1)--Higgs model with variable scalar 
field length seems to be achieved as outlined in (\ref{ContLim}).
The scaling of $\dts{\chi}{top}$ is still unclear and will be studied 
further. 
Also the systematic errors by the statistical uncertainties in the 
conditions (\ref{RenCond}) should be estimated.
The LCPs give strong evidence for a phase transition in 
$\kappa=\frac{1}{4}$ at $\beta=\infty$ and for a crossover for 
$\beta<\infty$.
%
%
%

%
\end{document}